**Aleksandr CARIOW** and Galina CARIOWA

WEST POMIERANIAN UNIVERSITY OF TECHNOLOGY, SZCZECIN
Żołnierska St. 49, 71-210, Szczecin
(corresponding author – **Aleksandr Cariow**)

# Some Schemes for Implementation of Arithmetic Operations with Complex Numbers Using Squaring Units

**Abstract**

In this paper, new schemes for a squarer, multiplier and divider of complex numbers are proposed. Traditional structural solutions for each of these operations require the presence some number of general-purpose binary multipliers. The advantage of our solutions is a removing of multiplications through replacing them by less costly squarers. We use Logan's trick and quarter square technique, which propose to replace the calculation of the product of two real numbers by summing the squares. Replacing usual multipliers on digital squares implies reducing power consumption as well as decreases hardware circuit complexity. The squarer requiring less area and power as compared to general-purpose multiplier, it is interesting to assess the use of squarers to implementation of complex arithmetic.

**Keywords**: complex number arithmetic, squaring unit, implementation complexity reduction, hardware implementation

## 1. Introduction

Today, mathematical operations with complex numbers are commonly used in numerous computer science applications, such as, digital signal and image processing, telecommunication, wireless data transmission, computer games and 3D machine graphics that require data processing in real time [1-4]. Many algorithms for computing convolutions, correlations, fast orthogonal transforms, etc. require complex-valued data processing. In today's processors it is quite common to have hardware implementation of all basic operations on real-valued operands. To our knowledge, there are no implementations of complex-valued squaring, multiplication and division at the hardware level on conventional processors. With a permanent increase in capacity of integrated circuits, it is timely to consider hardware-based implementation of an extended set of operations. In complex-valued arithmetic the most time and area consuming operations are squaring, multiplication and division because these operations contains real-valued multiplications and in the latter case – real-valued divisions. In turn, multiplication and division of two real numbers are also more time and area consuming operations than addition or subtraction of real numbers. What is more, the division is even more complicated and expensive than multiplication. Thus, it can be argued that the implementation complexity of these operations is quite high. It can be shown that the complexity of the hardware implementation of the listed operations can be reduced by replacing the multipliers by the squaring units. It should be noted that squares are a special case of multiplication where both operands are identical. For this reason designers often use general-purpose multipliers to implement the squaring units by connecting a multiplier's inputs together. Even though using general-purpose multipliers that are available as part of design packages reduces design time, it results in increased area and power requirements for the design [5]. Meanwhile, since the two operands are identical, some rationalizations can be made during the implementation of a dedicated squarer. In particular, unlike the general-purpose multiplier a dedicated squaring unit will have only one input, which allows to simplify the circuit. The article [6] shows that the dedicated squaring unit requires less than half whole amount of the logic gates as compared to the general-purpose multiplier. Dedicated squarer is area efficient, consumes less energy and dissipates less power as compared to general purpose multiplier. Proceeding from what has been said, we can conclude that the use of squaring units in the hardware implementation of complex-valued operations can be unusually effective.

Next, we will consider issues related to the hardware implementation of the main complex-valued operations using squaring units. Moreover, we will mean by default that we are talking about a completely parallel implementation of the proposed schemes, when each arithmetic operation is implemented by separate arithmetical unit.

## 2. Preliminaries

A complex number can be squared by multiplying by itself. Then in mathematical terms, such operation can be expressed as

$$z^2 = (a+jb)(a+jb) = (a^2 - b^2) + 2abj \qquad (1)$$

It is well known, that complex multiplication requires four real multiplications and two real additions, because:

$$(a_1 + jb_1)(a_2 + jb_2) = a_1a_2 - b_1b_2 + j(a_1b_2 + b_1a_2) \qquad (2)$$

So, we can observe that the direct computation of (2) requires four real multiplications and two real additions.

By 1805 Gauss had discovered a way of reducing the number of real multiplications to three. Then, it is possible to perform the complex multiplication with three real multiplication and five real additions, because [7]:

$$(a_1 + jb_1)(a_2 + jb_2) = a_1a_2 - b_1b_2 + \\ + j[(a_1 + b_1)(a_2 + b_2) - a_1a_2 - b_1b_2] \qquad (3)$$

To compute the quotient of two complex numbers, we multiply the numerator and denominator by the complex conjugate of the denominator:

$$z_3 = \frac{z_1}{z_2} = \frac{a_1 + jb_1}{a_2 + jb_2} = \frac{(a_1 + jb_1)(a_2 + jb_2)}{(a_2 + jb_2)(a_2 + jb_2)} = \\ = \frac{a_1a_2 - ja_1b_2 + ja_2b_1 - j^2b_1b_2}{a_2^2 + b_2^2} = \\ = \frac{a_1a_2 + b_1b_2}{a_2^2 + b_2^2} + \frac{a_2b_1 - a_1b_2}{a_2^2 + b_2^2}j \qquad (4)$$

In 1971, Logan noted that the multiplication of two real numbers can be performed using the following expression [8]:

$$ab = \frac{1}{2}[(a+b)^2 - a^2 - b^2] \qquad (4)$$

Another trick for multiplication of two real number is the so-called quarter square method [9]. The quarter square multiplication can be expressed as.

$$ab = \frac{1}{4}[(a+b)^2 - (a-b)^2] \qquad (5)$$

Applying these two methods and replacing the multiplication by calculating the squares, the hardware complexity of implementing squaring operations, multiplication and division of complex numbers can be reduced. In the next part of the article such an opportunity will be shown.

## 3. The schemes

Let we apply Logan's trick to the calculation of the square of a complex number.

Let $a_1 + jb_1$ is a complex number. Then the procedure for calculation of square of a complex number, represented in compact matrix notation, can be written as follows:

$$\mathbf{Y}_{2\times 1} = \mathbf{A}_2 \mathbf{T}_{2\times 3}[\mathbf{T}_{3\times 2}\mathbf{X}_{2\times 1}]^2 \qquad (6)$$

$$\mathbf{X}_{2\times 1} = [a_1, b_1]^T, \quad \mathbf{Y}_{2\times 1} = [c_1, d_1]^T,$$

$$\mathbf{T}_{3\times 2} = \begin{bmatrix} 1 & 0 \\ 0 & 1 \\ 1 & 1 \end{bmatrix}, \quad \mathbf{T}_{2\times 3} = \begin{bmatrix} 1 & 1 & 0 \\ 0 & 0 & 1 \end{bmatrix}, \quad \mathbf{A}_2 = \begin{bmatrix} 1 & 0 \\ -1 & 1 \end{bmatrix},$$

where $j = \sqrt{-1}$, symbol $[*]^2$ means squaring all the entries of the vector inscribed inside of the square brackets, $a_1$ - is real part of complex number, $b_1$ - is an imaginary part of complex number, $c_1$ - is real part of square of complex number, and $d_1$ - is an imaginary part of square of complex number.

Fig. 1 shows a schematic diagram for calculating the square of a complex number using squaring units. In this paper, all schemes are oriented from left to right. Straight lines in the schemes denote the operations of data transfer. The circles in these schemes show the operation of multiplication by a number inscribed inside a circle. The small squares in this figures indicate the squaring operations. Points where lines converge denote summation. We use the usual lines without arrows on purpose, so as not to clutter the scheme.

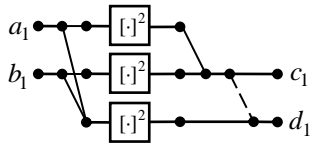

Fig. 1. Schematic diagram for calculating the square of a complex number using squaring units

Now we apply the Gauss's trick and the quarter square method to the calculation of the product of two complex numbers.

Let $a_1 + jb_1$ is a complex multiplier, $a_2 + jb_2$ is a complex multiplicand and $c_1 + jd_1$ is a complex number product. Then the computation procedure for complex number multiplication, represented in matrix notation, can be written as follows:

$$\mathbf{Y}_{2\times 1} = \mathbf{D}_2 \mathbf{A}_{2\times 3} \mathbf{A}_{3\times 6}[\widehat{\mathbf{H}}_6 \mathbf{A}_{6\times 4} \mathbf{X}_{4\times 1}]^2 \qquad (7)$$

$$\mathbf{A}_{6\times 4} = \begin{bmatrix} 0 & 0 & 1 & -1 \\ 1 & 0 & 0 & 0 \\ 0 & 0 & 1 & 1 \\ 0 & 1 & 0 & 0 \\ 0 & 0 & 0 & 1 \\ 1 & -1 & 0 & 0 \end{bmatrix}, \quad \mathbf{A}_{3\times 6} = \begin{bmatrix} 1 & -1 & 0 & 0 & 0 & 0 \\ 0 & 0 & 1 & -1 & 0 & 0 \\ 0 & 0 & 0 & 0 & 1 & -1 \end{bmatrix},$$

$$\widehat{\mathbf{H}}_6 = \mathbf{I}_3 \otimes \mathbf{H}_2 = \begin{bmatrix} 1 & 1 & & & & \\ 1 & -1 & & & & \\ & & 1 & 1 & & \\ & & 1 & -1 & & \\ & & & & 1 & 1 \\ & & & & 1 & -1 \end{bmatrix}, \quad \mathbf{I}_3 = \begin{bmatrix} 1 & & \\ & 1 & \\ & & 1 \end{bmatrix},$$

$$\mathbf{H}_2 = \begin{bmatrix} 1 & 1 \\ 1 & -1 \end{bmatrix}, \quad \mathbf{A}_{2\times 3} = \begin{bmatrix} 1 & 0 & 1 \\ 0 & 1 & 1 \end{bmatrix}, \quad \mathbf{D}_2 = diag(1/4, 1/4),$$

$$\mathbf{X}_{4\times 1} = [a_1, b_1, a_2, b_2]^T$$

where „$\otimes$" denotes the Kronecker product of two matrices [10], $a_1, a_2$ - are real parts of complex multiplier and multiplicand, $b_1, b_2$ - are imaginary parts of complex multiplier and multiplicand, and $c_2$ - is real part and $d_2$ is an imaginary part of complex number product respectively.

Fig. 2 shows a scheme for multiplying two complex numbers using squaring units. The rectangles indicate the operations of multiplication by the matrices $\mathbf{H}_2$.

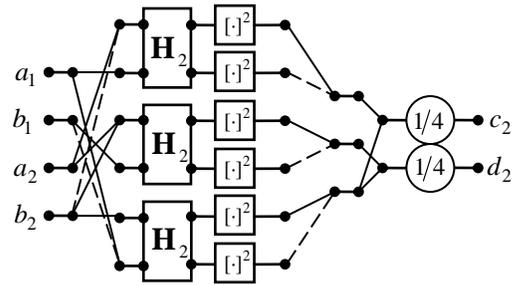

Fig. 2. Schematic diagram for complex-valued multiplication using squaring units

Finally we apply Logan's trick to the calculation of quotient of two complex numbers. Let $a_1 + jb_1$ is a complex dividend, $a_2 + jb_2$ is a complex divider and $c_3 + jd_3$ is a quotient of two complex numbers. Then the corresponding procedure, represented in matrix form, can be written as follows:

$$\mathbf{Y}_{2\times 1} = \widehat{\mathbf{D}}_2 \widehat{\mathbf{A}}_{2\times 3} \mathbf{A}_{3\times 4} \mathbf{A}_{4\times 8}[\mathbf{A}_8 \mathbf{P}_{8\times 4} \mathbf{X}_{4\times 1}]^2 \qquad (8)$$

$$\mathbf{X}_{4\times 1} = [a_1, a_2, b_1, b_2]^T, \quad \mathbf{P}_{8\times 4} = \mathbf{1}_2 \otimes \mathbf{I}_4 = \begin{bmatrix} 1 & & & \\ & 1 & & \\ & & 1 & \\ & & & 1 \\ \hline 1 & & & \\ & 1 & & \\ & & 1 & \\ & & & 1 \end{bmatrix},$$

$$\mathbf{A}_8 = \begin{bmatrix} 1 & 1 & 0 & 0 & & & & \\ 0 & 0 & 1 & 1 & & \mathbf{0}_4 & & \\ 0 & 1 & 1 & 0 & & & & \\ 1 & 0 & 0 & 1 & & & & \\ \hline & & & & 1 & 0 & 0 & 0 \\ & \mathbf{0}_4 & & & 0 & 0 & 1 & 0 \\ & & & & 0 & 1 & 0 & 0 \\ & & & & 0 & 0 & 0 & 1 \end{bmatrix},$$

$$\mathbf{A}_{4\times 8} = \begin{bmatrix} 1 & 1 & 0 & 0 & 0 & 0 & 0 & 0 \\ 0 & 0 & 1 & -1 & 0 & 0 & 0 & 0 \\ 0 & 0 & 0 & 0 & 1 & 1 & 0 & 0 \\ 0 & 0 & 0 & 0 & 0 & 0 & 1 & 1 \end{bmatrix},$$

$$\mathbf{A}_{3\times 4} = \begin{bmatrix} 1 & 0 & 0 & 0 \\ 0 & 1 & 0 & 0 \\ 0 & 0 & 1 & 1 \end{bmatrix}, \hat{\mathbf{A}}_{2\times 3} = \begin{bmatrix} 1 & 0 & -1 \\ 0 & 1 & -1 \end{bmatrix},$$

$$\hat{\mathbf{D}}_2 = \mathbf{I}_2 \otimes 1/2(a_2^2 + b_2^2)$$

and $\mathbf{0}_{N\times M}$ is an $M \times N$ matrix of zeros (a matrix where every element is equal to zero), $\mathbf{I}_N$ - is an identity matrix, $\mathbf{1}_2 = [1,1]^T$ [11].

Fig. 3 shows a schematic diagram for calculation of quotient of two complex numbers using squaring units.

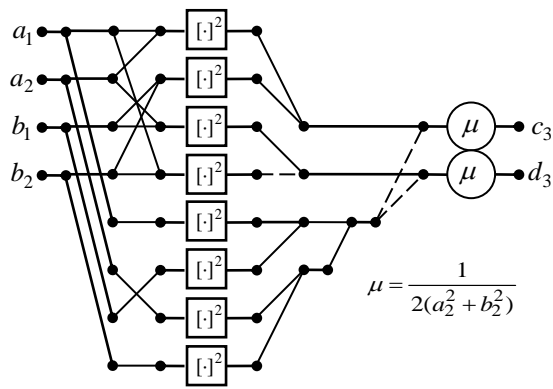

Fig. 3. Schematic diagram for complex-valued division using squaring units

## 4. Implementation complexity

Tables 1 and 2 show the nomenclature and the number of arithmetic units that are necessary for the hardware implementation of the described complex-valued operations in the case of using the traditional arithmetic units and in the case of using the squaring units, respectively. Analyzing these data, it is easy to verify that in some cases the hardware implementation of the proposed schemes is more efficient than the hardware implementation using traditional arithmetic blocks.

Tab. 1. The nomenclature and the number of arithmetic units necessary for the implementation of complex-valued operations using direct methods

| Type of arithmetic unit | Type of implemented complex-valued operation | | |
|---|---|---|---|
| | squaring | multiplication | division |
| adders | 1 | 2 | 3 |
| squarers | 2 | — | 2 |
| multipliers | 1 | 4 | 4 |
| dividers | — | — | 2 |

Tab. 2. The nomenclature and the number of arithmetic units necessary for the implementation of complex-valued operations using proposed schemes

| Type of arithmetic unit | Type of implemented complex-valued operation | | |
|---|---|---|---|
| | squaring | multiplication | division |
| adders | 3 | 14 | 11 |
| squarers | 3 | 6 | 8 |
| multipliers | — | — | — |
| dividers | — | — | 2 |

## 5. Conclusion

The article presents three new schemes for fully parallel hardware implementation of basic complex arithmetic operations, namely: squaring, multiplication and division using squaring units. To reduce the hardware complexity (number of two-operand multipliers), we exploit the Logan's identity and quarter square method for replacing the binary multipliers by the squaring units. If the requirements for the speed of solving the task are not high, then in order to minimize the chip area and reducing power consumption, only one squaring unit can be reused many times. We provide an opportunity for an inquisitive reader to make sure of the effectiveness of the proposed solutions independently.

_________________________________________________

**Prof., DSc, PhD Aleksandr CARIOW**

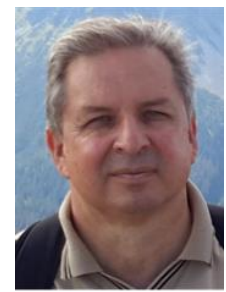

He received the Candidate of Sciences (PhD) and Doctor of Sciences degree (DSc or Habilitation) in Computer Sciences from LITMO of St. Petersburg, Russia in 1984 and 2001, respectively. In September 1999, he joined the faculty of Computer Sciences at the West Pomeranian University of Technology, Szczecin, Poland, where he is currently a professor and chair of the Department of Computer Architectures and Telecommunications. His research interests include digital signal processing algorithms, VLSI architectures, and data processing parallelization.

*e-mail: acariow@wi.zut.edu.pl*

**PhD Galina CARIOWA**

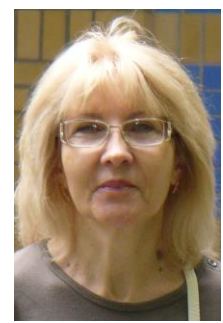

She received the MSc degree in Mathematics from Moldavian State University, Chişinău in 1976 and PhD degree in computer science from West Pomeranian University of Technology, Szczecin, Poland in 2007. She is currently working as an assistant professor of the Department of Multimedia Systems. She is also an Associate-Editor of World Research Journal of Transactions on Algorithms. Her scientific interests include numerical linear algebra and digital signal processing algorithms, VLSI architectures, and data processing parallelization.

*e-mail: gcariowa@wi.zut.edu.pl*